\documentclass{llncs}
\usepackage[utf8]{inputenc}
\usepackage[T1]{fontenc}
\usepackage{amsmath,stmaryrd,mathpartir}
\usepackage{amssymb}
\usepackage{proof}
\usepackage[]{graphicx}
\usepackage{paralist}
\usepackage{enumitem}
\usepackage{color}
\usepackage{xspace}
\usepackage{url}
\usepackage{bussproofs}
\usepackage{appendix}
\usepackage[inference]{semantic}
\usepackage{fancyhdr}
\usepackage{epsfig}
\usepackage{latexsym}
\usepackage{lmodern}
\usepackage{microtype}
\usepackage{lastpage} 
\usepackage{hyperref}
\usepackage{tabularx}
\usepackage{multirow}
\usepackage{listings}
\lstdefinestyle{customhaskell}{
    language=Haskell,
    showstringspaces=false,
    basicstyle=\footnotesize\rmfamily,
    keywordstyle=\bfseries\color{green!40!black},
    commentstyle=\itshape\color{blue!40!black},
    identifierstyle=\color{black},
    stringstyle=\color{orange!40!black},
}
\usepackage[dvipsnames*,svgnames]{xcolor}
\hypersetup{colorlinks=true,allcolors=DarkBlue,linkcolor=black}

\newcommand{\setof}[1]{\{ #1 \}}
\newcommand{\ap}[2]{\mbox{$\mathit{#1}(#2)$}}
\newcommand{\bap}[3]{\mbox{$\mathit{#1}(#2,#3)$}}
\newcommand{\tap}[4]{\mbox{$\mathit{#1}(#2,#3,#4)$}}

\newcommand{\step}[1]{\mathbin{\lower0.55ex\hbox{$\lhook\joinrel\xrightarrow{#1}$}}}
\newcommand{\semstep}[1]{\step{#1}}

\newcommand{\Activity}{\mathsf{Activity}}

\newcommand{\ContProv}{\mathsf{ContProv}}

\newcommand{\Comp}{\mathsf{Comp}}

\newcommand{\CompInstance}{\mathsf{CompInstance}}

\newcommand{\iComp}{\mathsf{iComp}}
\newcommand{\Val}{\mathsf{Val}}
\newcommand{\PermId}{\mathsf{PermId}}
\newcommand{\PermLvl}{\mathsf{PermLvl}}
\newcommand{\Perm}{\mathsf{Perm}}

\newcommand{\PermGrp}{\mathsf{PermGroup}}

\newcommand{\Uri}{\mathsf{Uri}}

\newcommand{\Manifest}{\mathsf{Manifest}}

\newcommand{\AppId}{\mathsf{AppId}}
\newcommand{\Cert}{\mathsf{Cert}}
\newcommand{\Res}{\mathsf{Res}}
\newcommand{\ResVal}{\mathsf{Val}}

\newcommand{\App}{\mathsf{App}}
\newcommand{\SysApp}{\mathsf{SysImgApp}}
\newcommand{\Intent}{\mathsf{Intent}}
\newcommand{\InstApps}{\mathsf{InstApps}}
\newcommand{\GrantedGroups}{\mathsf{PermsGr}}
\newcommand{\AppsPerms}{\mathsf{AppPS}}
\newcommand{\AppsDefPerms}{\mathsf{AppDefPS}}
\newcommand{\CompInsRunning}{\mathsf{CompInsRun}}
\newcommand{\OpType}{\mathsf{OpTy}}

\newcommand{\DelPPerms}{\mathsf{DelPPerms}}

\newcommand{\DelTPerms}{\mathsf{DelTPerms}}

\newcommand{\AppsResCont}{\mathsf{ARVS}}
\newcommand{\SentIntents}{\mathsf{Intents}}
\newcommand{\AppsCert}{\mathsf{Certs}}
\newcommand{\AppsManifest}{\mathsf{Manifests}}
\newcommand{\ImageApps}{\mathsf{SysImage}}
\newcommand{\AndroidState}{\mathsf{AndroidST}}

\newcommand{\Action}{\mathsf{Action}}
\newlength{\bcextramargin}
\newenvironment{changemargin}[2]{\begin{list}{}{%
\setlength{\topsep}{0pt}%
\setlength{\leftmargin}{0pt}%
\setlength{\rightmargin}{0pt}%
\setlength{\listparindent}{\parindent}%
\setlength{\itemindent}{\parindent}%
\setlength{\parsep}{0pt plus 1pt}%
\addtolength{\leftmargin}{#1}%
\addtolength{\rightmargin}{#2}%
}\item }{\end{list}} 

\newcommand{\actdefsection}[1]{
\begin{changemargin}{-\bcextramargin}{0pt}
\vspace{1ex}
\noindent
\textbf{{#1}}
\end{changemargin}
}

\newenvironment{absolutelynopagebreak}
  {\par\nobreak\vfil\penalty0\vfilneg
   \vtop\bgroup}
  {\par\xdef\tpd{\the\prevdepth}\egroup
   \prevdepth=\tpd}

\newcommand{\Mathexecrel}[3]{#1 \semstep{#2} #3}
\newtheorem{prop}{Property}
\newcommand{\eqdef}{\stackrel{{\rm def}}{=}}
\newcommand{\figref}[1]{Figure~\ref{#1}}

\newcommand{\lemref}[1]{Lemma~\ref{#1}}
\begin{document}
\pagestyle{headings}
\bibliographystyle{plain}
\title{A certified reference validation mechanism for the permission model of Android
\thanks{Partially funded by project ANII-FCE\_1\_2014\_1\_103803: Mecanismos aut\'onomos de seguridad certificados para sistemas computacionales m\'oviles, Uruguay.}
} \author{Gustavo Betarte\inst{1} \and Juan Campo\inst{1} \and Felipe Gorostiaga\inst{2} \and Carlos
    Luna\inst{1}} \institute{InCo, Facultad de Ingenier\'ia, Universidad de la Rep\'ublica, Uruguay. 
   \\ \email{\{gustun,jdcampo,cluna\}@fing.edu.uy} 
\and IMDEA Software Institute, Spain.
 \\ \email{felipe.gorostiaga@imdea.org}
}

\maketitle

\begin{abstract}                                          
Android embodies security mechanisms at both OS and application level. In this platform application security is built primarily upon a system of permissions which specify restrictions on the operations a particular process can perform.
The critical role of these security mechanisms makes them a prime target for (formal) verification. We present an idealized model of a reference monitor  of the novel mechanisms of Android 6 (and further), where it is possible to grant permissions at run time. Using the programming language of the proof-assistant \texttt{Coq} we have developed a functional implementation of the reference validation mechanism and certified its correctness with respect to the specified reference monitor. Several properties concerning the permission model of Android 6 and its security mechanisms have been formally formulated and proved. Applying the program extraction mechanism provided by \texttt{Coq} we have also derived a certified \texttt{Haskell} prototype of the reference validation mechanism.
\end{abstract}
\section{Introduction}
\label{sec:intro}
The Android \cite{AndroidProy} platform for mobile devices captures more than 85\%
of the total market-share \cite{reporteGartner}. Mobile devices allow people to develop multiple tasks in different areas, regrettably, the benefits of using them are counteracted by increasing security risks.

Android embodies security mechanisms at both OS and application level. Application security is built primarily upon a system of permissions, which specify restrictions on the operations a particular process can perform. Permissions in Android are basically tags that developers declare in their applications, more precisely in the so-called application \emph{manifest}, to gain access to sensitive resources. On all versions of Android an application must declare both the normal and the dangerous permissions it needs in its application manifest. However, the effect of that declaration is different depending on the system version and the application's target SDK level \cite{android6}.
In particular, if a device is running Android 6 (Marshmallow) and the application's target SDK is 23 or higher the application has to list the permissions in the manifest, and it must request each dangerous permission it needs while the application is running. The user can grant or deny each permission, and the application can continue to run with limited capabilities even if the user denies a permission request. This modification of the access control and decision process, on the one side, streamlines the application install process, since the user does not need to grant permissions when he/she installs or updates an application. On the other hand, as users can revoke the (previously granted) permissions at any time, the application needs to check whether it has the corresponding privileges every time it attempts to access a resource on the device \cite{android6}.
The important and critical role of these security mechanisms makes them a prime target for (formal) verification.

Security models play an important role in the design and evaluation of security mechanisms of systems. Their importance was already pointed out in 1972 in the Anderson report \cite{Anderson:1972}, where the concept of \emph{reference monitor} was first introduced. This concept defines the design requirements for implementing what is called a \emph{reference validation mechanism}, which shall be responsible for enforcing the access control policy of a system.
The work presented here is concerned with the formal analysis and verification of properties performed on an idealized model that abstracts away the specifics of any particular implementation, and yet provides a realistic setting in which to explore the issues that pertain to the realm of (critical) security mechanisms of Android.

 \paragraph{Contributions}
 In \cite{DBLP:conf/ictac/BetarteCLR15,BetarteCLR16}, we have presented a 
formal specification of an idealized formulation of the permission model of version 5 of Android.
Here we present an enriched version of that model which can be used to perform a formal analysis of the novel mechanisms of Android 6, which make it possible to grant permissions at run time.
Furthermore, using the programming language of \texttt{Coq} \cite{coq-manual} we have developed an executable (functional) specification of the reference validation mechanism and it has been proved that those functions conform to the axiomatic specification as specified in the model. Additionally, and using the program extraction mechanism provided by \texttt{Coq}, we have derived a certified \texttt{Haskell} prototype of the reference validation mechanism.
Several properties concerning the security model of Android 6 have been formally formulated and proved.

\paragraph{Organization of the paper}
Section \ref{sec:background} reviews the security mechanisms of Android. Section \ref{sec:model} describes the formal axiomatic specification of the Android security system and discusses some of the verified properties.
Section~\ref{sec:excspec} presents a functional (operational) semantics of the specified reference monitor and outlines its proof of correctness. This section also discusses security properties satisfied by the certified implementation of the security mechanisms.
Section~\ref{sec:relwork} considers related work and finally, Section \ref{sec:conclusion} concludes with a summary of our contributions and directions for future work.
The full formalization may be obtained from \cite{AndroidCoq:2016} and verified using the \texttt{Coq} proof assistant. 
\section{Android's security model}
\label{sec:background}
The architecture of Android takes the form of a software stack which comprises an OS, a run-time environment, middleware, services and libraries, and applications. An Android application is built up from \textit{components}.  A component is a basic unit that provides a particular functionality and that can be run by any other application with the right permissions. There exist four types of components \cite{fundamentals}: 
\begin{inparaenum}[i)]
\item \textit{activity}, which is essentially a user interface of the application; 
\item \textit{service},  a component that executes in background without providing an interface to the user;
\item \textit{content provider},  a component intended to share information among applications; and 
\item \textit{broadcast receiver}, a component whose objective is to receive messages, sent either by the system or an application, and trigger the corresponding actions. 
\end{inparaenum}
Activities, services and broadcast receivers are activated by a special kind of message called \textit{intent}. An intent makes it possible for different components to interact at runtime.
A intent filter specifies the types of intents that a component receiver can respond to \cite{fundamentals}. 
Applications usually need to use system resources to execute properly.
Since applications run inside sandboxes, this entails the existence of
a decision procedure (a reference validation mechanism) that guarantees the
authorized access to those resources. Decisions are made by following
security policies using a simple notion of permission. Every permission is identified by a name/text, has a protection level and may belong to a permission group.
There exist two principal classes of permissions: the ones defined by the application,
for the sake of self-protection; and those predefined by Android,
which are intended to protect access to resources and services of the
system. An application declares --in a XML file called \texttt{AndroidManifest}-- 
  the set of permissions it needs to acquire further capacities than the default ones. When an action involving permissions is required, the system determines which permissions every application has and either allows or denies its execution.

Depending on the protection level of the permission, the system defines the corresponding decision
procedure~\cite{protectionLevel}. There are four classes of permission levels:
\begin{inparaenum}[i)] 
\item \textit{Normal}, assigned to low risk permissions that grant access to isolated characteristics;
\item \textit{Dangerous}, permissions of this level are those that provide access to private data or control over the device. From version 6 of Android dangerous permissions are not granted at installation time;
\item \textit{Signature}, a permission of this level is granted
  only if the application that requires it and the application that
  defined it are both signed with the same certificate; and
\item \textit{Signature/System}, this level is assigned to permissions
  that regulate the access to critical system resources or services.
\end{inparaenum}
Additionally, an application can also declare the permissions that are
needed to access it. 
A running application may ask the user to grant it dangerous permission groups and ungrouped
permissions, who in turn can accept or decline this request.

If the execution of an action requires for an application to have certain permission the system will first make sure that this holds by means of the following rules:
\begin{inparaenum}[i)] 
\item the application must declare the permission as used in its manifest;
\item if the permission is of level \textit{Normal}, then the application does have it;
\item if the permission is of level \textit{Dangerous} and belongs to a permission group, such group must have been granted to the application;
\item if the permission is of level \textit{Dangerous} but is ungrouped, then it must have been individually granted to the application;
\item if the permission is of level \textit{Signature}, then the both the involved application and the one that declares it must have been signed with the same certificate;
\item lastly, if the permission is of level \textit{Signature/System}, then the involved application must have been signed with either the same certificate as the one who declares it or the certificate of the device manufacturer.
\end{inparaenum}
Otherwise, an error is thrown and the action is not executed.

Android provides two mechanisms by which an application can delegate
its own permissions to another one. These mechanisms are called
\textit{pending intents} and \textit{URI permissions}.
An intent may be defined by a developer to perform a particular
action. A \texttt{PendingIntent} is
an object which is associated to the action, a reference that might be
used by another application to execute that action. 
The \textit{URI permissions} mechanism can be used by an application
that has read/write access to a \textit{content provider} to
 delegate those permissions to another application. 
\section{Formalization of the permission model}
\label{sec:model}
In this section we provide a short account of the axiomatic semantics of the Android security system and discuss some of the verified security properties.

\paragraph {Formal language used} 
The \texttt{Coq} proof assistant provides a (dependently typed) functional programming language and a reasoning framework based on higher order logic to perform proofs of (complex) specifications and programs. \texttt{Coq} allows developing mathematical facts. This includes defining objects (sets, lists, streams, functions, programs); making statements (using basic predicates, logical connectives and quantifiers); and finally writing proofs. The type of propositions is called \begin{small}\verb+Prop+\end{small}.
The \texttt{Coq} environment provides program extraction towards languages like Ocaml and Haskell for execution of (certified) algorithms \cite{letouzey04,conf/types/Letouzey02}.

In this work, enumerated types and sum types are defined using Haskell-like notation; for example, $option\ T \eqdef None \mid Some\ (t:T)$. Record types are of the form $\left\{ l_1 : T_1,\ldots, l_n : T_n \right\}$,
whereas their elements are of the form $\{t_1, \ldots, t_n \}$. Field selection is written as $r.l_i$.
We also use $\setof{T}$ to denote the set of elements of type $T$. Finally, the symbol $\times$ defines tuples, and 
$nat$ is the datatype of natural numbers. 
We omit \texttt{Coq} code for reasons of clarity; this code is avalilable in \cite{AndroidCoq:2016}. 

\subsection{Model states}
\label{sec:model:stateI}

The Android security model we have developed has been formalized as an abstract state machine. In this model, states ($\AndroidState$) are modelled as 12-tuples that respectively store data about the applications installed, their permissions and the running instances of components; the formal definition appears in Figure~\ref{fig:model:state}.

\begin{figure}
\scriptsize
\begin{displaymath}
\begin{array}[t]{l@{\,}l@{\,\,}l}
\OpType & ::= & read\; | \; write\; |  \; rw  \\

\PermLvl & ::= & dangerous\; | \; normal\; | \; signature\; | \; Signature/System	\\

\Perm & ::= & \PermId \times option~\PermGrp \times \PermLvl \\

\InstApps & ::= & \setof{\AppId} \\
\GrantedGroups & ::= & \setof{\AppId \times \PermGrp} \\ 
\AppsPerms & ::= & \setof{\AppId \times \PermId} \\ 
\CompInstance & ::= & \iComp \times \Comp \\
\CompInsRunning & ::= & \setof{\CompInstance} \\
\DelPPerms & ::= & \setof{\AppId \times \ContProv \times \Uri \times \OpType} \\ 
\DelTPerms & ::= & \setof{\iComp \times \ContProv \times \Uri \times \OpType} \\
\AppsResCont & ::= & \setof{\AppId \times \Res \times \ResVal} \\
\SentIntents & ::= & \setof{\iComp \times \Intent} \\ 
\Manifest & ::= & \setof{\Comp} \times option~nat \times option~nat~ \times 
 \setof{\PermId} \times \setof{\Perm} \times option~\PermId \\
\AppsManifest & ::= & \setof{\AppId \times \Manifest} \\ 
\AppsCert & ::= & \setof{\AppId \times \Cert} \\
\AppsDefPerms & ::= & \setof{\AppId \times \Perm} \\ 
\ImageApps & ::= & \setof{\SysApp}  \\ [1ex]
\AndroidState & ::= &  \InstApps \times \GrantedGroups \times \AppsPerms \times  \CompInsRunning \times \DelPPerms \times \DelTPerms\ \times \\
                    &&  \AppsResCont \times \SentIntents~ \times \AppsManifest \times \AppsCert \times \AppsDefPerms \times \ImageApps \\
\end{array}
\end{displaymath}
\caption{Android state}
\label{fig:model:state}
\end{figure}

The type $\PermId$ represents the set of permissions identifiers;
$\PermGrp$, the set of permission groups identifiers;
$\Comp$, the application components whose code will run on the system;
$\AppId$ represents the set of application identifiers;
$\iComp$ is the set of identifiers of running instances of application components;
$\ContProv$ is a subset of $\Comp$, a special type of component that allows sharing resources among different applications;
a member of the type $\Uri$ is a particular uri (uniform resources identifier);
the type $\Res$ represents the set of resources an application can have (through its $content~providers$, members of $\ContProv$);
the type $\Val$ is the set of possible values that can be written on resources;
an intent --i.e. a member of type $\Intent$-- represents the intention of a running component instance to start or communicate with other applications;
a member of $\SysApp$ is a special kind of applications which are deployed along with the OS itself and has certain privileges, like being impossible to uninstall. 

The first component of a state records the identifiers ($\AppId$) of the applications installed by the user. The second and third components of the state keep track, respectively, of the permission groups ($\PermGrp$) and ungrouped permissions granted to each application present in the system, both the ones installed by the user and the system applications.
The fourth component of the state stores the set of running component instances ($\CompInstance$), while the components $\DelPPerms$ and $\DelTPerms$ store the information concerning permanent and temporary permissions delegations, respectively\footnote{A
permanent delegated permission represents that
an app has delegated permission to perform an operation on the resource
identified by an URI. A temporary
delegated permission refers to a permission that has been
delegated to a component instance.}.
The seventh and eight components of the state store respectively the values ($\Val$) of resources ($\Res$) of applications and the set of intents ($\Intent$) sent by running instances of components ($\iComp$) not yet processed.
The four last components of the state record information that represents the manifests of the applications installed by the user, the certificates ($\Cert$) with which they were signed and the set of permissions they define. The last component of the state stores the set of (native) applications installed in the Android system image, information that is relevant when granting permissions of level $Signature/System$.
A manifest ($\Manifest$) is modelled as a 6-tuple that respectively declare application components (set of components, of type $\Comp$, included in the application); optionally, the minimum version of the Android SDK required to run the application; optionally,  the version of the Android SDK targeted on development; the set of permissions it may need to run at its maximum capability; the set of permissions it declares; and the permission required to interact with its components, if any. 
Application components are all denoted by a component identifier. 
A content provider ($\ContProv$),  in addition, encompasses a mapping to the managed resources from the URIs assigned to them for external access. 
While the components constitute the static building blocks of an application, all runtime operations are initiated by component instances, which are represented in our model as members of an abstract type. 

We define a notion of valid state, through the predicate $valid\_state$ on the elements of type $\AndroidState$, that captures several well-formedness conditions. The definition is provided in Appendix~\ref{app:validstate}.

\subsection{Action semantics}
\label{sec:model:actsemI}
The axiomatic semantics of the Android security system is modeled by defining a set of actions,
and providing their semantics as state transformers.
Table \ref{table:actions} summarises a subset of the actions specified in our model, which provide coverage to the different functionalities of the Android security model. 

\begin{table}[thb!]
\scriptsize
\centering
\begin{tabularx}{\linewidth}{|l X|}
	\hline
	$\mathtt{install}~app~m~c~lRes$	& Install application with id $app$, whose manifest is $m$, is signed with certificate $c$ and its resources list is $lRes$. \\
	\hline
	$\mathtt{uninstall}~app$	& Uninstall the application with id $app$. \\
	\hline
	$\mathtt{grant}~p~app$	& Grant the permission $p$ to the application $app$. \\ 
	\hline
	$\mathtt{revoke}~p~app$	& Remove the permission $p$ from the application $app$. \\
	\hline
	$\mathtt{grantPermGroup}~g~app$	& Grant the permission group $g$ to the application $app$. \\
	\hline
	$\mathtt{revokePermGroup}~g~app$	& Remove the permission group $g$ from the application $app$. \\
	\hline
	$\mathtt{hasPermission}~p~app$	& Check if the application $app$ has the permission $p$. \\
	\hline
	$\mathtt{read}~ic~cp~u$	& The running comp. $ic$ reads the resource corresponding to URI $u$ from content provider $cp$. \\
	\hline
	$\mathtt{write}~ic~cp~u~val$	& The running comp. $ic$ writes value $val$ on the resource corresponding to URI $u$ from content provider $cp$. \\
	\hline
	$\mathtt{startActivity}~i~ic$	& The running comp. $ic$ asks to start an activity specified by the intent $i$. \\
	\hline
	$\mathtt{startActivityRes}~i~n~ic$	& The running comp. $ic$ asks to start an activity specified by the intent $i$, and expects as return a token $n$. \\
	\hline
	$\mathtt{startService}~i~ic$	& The running comp. $ic$ asks to start a service specified by the intent $i$. \\
	\hline
	$\mathtt{sendBroadcast}~i~ic~p$	& The running comp. $ic$ sends the intent $i$ as broadcast, specifying that only those components who have the permission $p$ can receive it. \\
	\hline
	$\mathtt{sendOrdBroadcast}~i~ic~p$	& The running comp. $ic$ sends the intent $i$ as an ordered broadcast, specifying that only those components who have the permission $p$ can receive it. \\
	\hline
	$\mathtt{sendSBroadcast}~i~ic$	& The running comp. $ic$ sends the intent $i$ as a sticky broadcast. \\
	\hline
	$\mathtt{resolveIntent}~i~app$	& Application $app$ makes the intent $i$ explicit. \\
	\hline
	$\mathtt{receiveIntent}~i~ic~app$	& Application $app$ receives the intent $i$, sent by the running comp. $ic$. \\
	\hline
	$\mathtt{stop}~ic$	& The running comp. $ic$ finishes its execution. \\
	\hline
	$\mathtt{grantP}~ic~cp~app~u~pt$	& The running comp. $ic$ delegates permanent permissions to application $app$. This delegation enables $app$ to perform operation $pt$ on the resource assigned to URI $u$ from content provider $cp$. \\
	\hline
	$\mathtt{revokeDel}~ic~cp~u~pt$	& The running comp. $ic$ revokes delegated permissions on URI $u$ from content provider $cp$ to perform operation $pt$. \\
	\hline
	$\mathtt{call}~ic~sac$	& The running comp. $ic$ makes the API call $sac$. \\
	\hline
  \end{tabularx}
  \caption{Actions}
  \label{table:actions}
\end{table}

The behaviour of actions (of type $\Action$) is specified by a
precondition $Pre$ and by a postcondition $Post$ of respective types:
$Pre  :  \AndroidState \rightarrow \Action \rightarrow Prop$, and $ Post  :  \AndroidState \rightarrow \Action \rightarrow \AndroidState \rightarrow Prop$.
For instance, the axiomatic semantics of the \texttt{install} action is given by:\\ \\
\footnotesize
$\begin{array}{l}
\quad \quad Pre(s,\texttt{install}\ app\ m\ c\ lRes) \eqdef\\ 
\quad \quad \quad \quad \neg isAppInstalled(app,s)\ \land\
\neg has\_duplicates\_cmp(m)\ \land \\
\quad \quad \quad \quad \forall c:\Comp, c \in cmp(m)\ \rightarrow\  c \notin cmpInState(s)\ \land \\
\quad \quad \quad \quad \neg hasDuplicates\_perm(m)\  \land\
authPerms(m,s)\ \land \\
\quad \quad \quad \quad \forall c:\Comp, c \in cmp(m)\ \rightarrow\ cmpDeclareIntentFilterCorrectly(c) \\ 
\quad \quad Post(s,\texttt{install}\ app\ m\ c\ lRes, s') \eqdef\\ 
\quad \quad \quad \quad addManifest(m,app,s,s')\ \land\
addCert(c,app,s,s')\ \land \\
\quad \quad \quad \quad addDefPerms(app,m,s,s')\ \land\
addApp(app,s,s')\ \land \\
\quad \quad \quad \quad addRes(app,lRes,s,s')\ \land\
initializePermLists(app,s,s')\ \land \\
\quad \quad \quad \quad sameOtherFields\_install(s,s')\
\end{array}$ \\ 
\normalsize 

\noindent The precondition of action $\texttt{install}\ app\ m\ c\ lRes$ in a state $s$ requires that the application is not already installed, and that no system application has its same identifier ($\neg isAppInstalled(app,s)$).
The identifiers of the components listed in its manifest ($c \in cmp(m)$) must be different from each other ($\neg has\_duplicates\_$\\$cmp(m)$), and also different from those of the components already present in the device ($c \notin cmpInState(s)$).
The permissions defined by the application to be installed must be different from each other ($\neg hasDuplicates\_perm(m)$) as well as different from those defined by other applications ($authPerms(m,s)$).
Finally, the intent filters of its components ($c \in cmp(m)$) must be well defined; the specified types of intents must match ($cmpDeclareIntentFilterCorrectly(c)$).

The postcondition of action $\texttt{install}\ app\ m\ c\ lRes$, with initial state $s$ and final state $s'$, specifies that its manifest ($m$), the certificate with which it was signed ($c$) and the permissions it defines that are not system permissions must be added to the state bound to the application identifier (predicates $addManifest,addCert,$ and $addDefPerms$); which in turn must be included in the list of installed applications ($addApp(app,s,s')$).
The application resources are initialized with the initial value $initVal$ ($addRes(app,lRes,s,s')$), while the lists of permission and permission groups granted to the application are initialized as empty ($initializePermLists(app,s,s')$).
The rest of the system components remain unchanged ($sameOtherFields\_install(s,s')$).

\subsection{Executions}
\label{sec:error}
There can be attempts to execute an action on a state that does not verify the precondition of that action. In the presence of one such situation
the system answers with a corresponding error code (of type $ErrorCode$). 

Executing an action $a$ over a valid state $s$ produces a new state $s'$ and a corresponding answer $r$ (denoted $s\step{a/r}s'$),
where the relation between the former state and the new one is given by the postcondition relation $Post$.

\begin{displaymath}
\footnotesize
\begin{array}{c}
\inference[]{$$valid\_state(s)$$ \hspace{.2cm} $$Pre(s, a)$$ \hspace{.2cm} $$Post(s, a, s')$$}{$$s\step{a/ok}s'$$} 
\hspace{0.5cm} 
\inference[]{$$valid\_state(s)$$ \hspace{.2cm} $$ErrorMsg(s, a, ec)$$}{$$s\step{a/error(ec)}s$$}
\end{array}
\end{displaymath}
\noindent Whenever an action occurs for which the precondition holds, the valid state may change in such a way that the action postcondition is established. The notation $s \step{a/ok} s'$ may be read as \textit{the execution of the action $a$ in a valid state $s$ results in a new state $s'$}. However, if the precondition is not satisfied, then the valid state $s$ remains unchanged and the system answer is the error message determined by a relation $ErrorMsg$\footnote{Given a state $s$, an action $a$ and an error code $ec$, $\tap{ErrorMsg}{s}{a}{ec}$ holds iff $error~ec$ is an acceptable response when the execution of $a$ is requested on state $s$.}. Formally, the possible answers of the system are defined by the type $\textit{Response} \eqdef ok\ |\ error\ (ec : ErrorCode)$, where  $ok$ is the answer resulting from a successful execution of an action.
One-step execution with error management preserves valid states.
\begin{lemma} [Validity is invariant]
\footnotesize
\label{lemma:valid-state-correct}
\mbox{}\\
$\begin{array}{l}
\forall\ (s\ s':\AndroidState)(a:\Action) (r:Response),
s\step{a/r}s' \rightarrow valid\_state(s')
\end{array}$
\end{lemma}
The results presented in this work are obtained from valid states of the system.

\subsection{Reasoning over the specified reference monitor}
\label{sec:properties}
In this section we present and discuss some relevant properties that can be established concerning the Android Marshmallow security framework. In particular we shall focus on vulnerabilities that if exploited would allow violations to the intended security policy.
The helper functions and predicates used to define the properties and lemmas discussed in this paper are presented and described in Table \ref{tab:model:semantic_descriptions}. 
The full formal definition of the lemmas as well as those of other security properties that were formally analyzed can be found in \cite{AndroidCoq:2016}, along with their corresponding proofs. We also include an informal description of each property, in italics.
%
\begin{table}[thb!]
\scriptsize
\centering
\begin{tabularx}{\linewidth}{|l X|}
\hline
\textbf{Function/Predicate} & \textbf{Description} \\
\hline
$appHasPermission(app,p,s)$ & holds iff $app$ is considered to have permission $p$ on state $s$. \\
\hline
$canGrant(cp,u,s)$	& holds iff the content provider $cp$ allows the delegation of permissions over the resource at URI $u$ on state $s$. \\
\hline
$\tap{\mathit{canStart}}{c'}{c}{s}$ & holds if the app containing component $c'$ (installed in $s$) has the required permissions to create a new running instance of $c$. \\
\hline
$cmpProtectedByPerm(c)$	& returns the permission by which the component $c$ is protected. \\
\hline
$componentIsExported(c)$	& holds iff the component $c$ is exported and can be accessed from other applications. \\
\hline
$existsRes(cp,u,s)$	& holds iff the URI $u$ belongs to the content provider $cp$ on $s$. \\
\hline
$getAppFromCmp(c,s)$	& given a component $c$ on $s$, returns the app to which it belongs. \\
\hline
$getAppRequestedPerms(m)$ & given the manifest $m$ of an app, returns the set of permissions it uses. \\
\hline
$getDefPermsForApp(app,s)$	& returns the set of permissions defined by $app$ on state $s$. \\
\hline
$getGrantedPermsForApp(app,s)$	& returns the set of indvidual permissions granted to $app$ on $s$. \\
\hline
$getInstalledApps(s)$ & returns the set of identifiers of the applications installed on $s$.\\
\hline
$getManifestForApp(app,s)$	& returns the manifest of application $app$ on state $s$. \\
\hline
$getPermissionId(p)$	& returns the identifier of permission $p$. \\
\hline
$getPermissionLevel(p)$	& returns the permission level of permission $p$. \\
\hline
$getRunningComponents(s)$	& returns the set of pairs consisting of a running instance id, and its associated component currently running on state $s$. \\
\hline
$inApp(c,app,s)$	& holds iff the component $c$ belongs to application $app$ on state $s$. \\
\hline
$permissionIsGrouped(p)$	& holds iff permission $p$ belongs to any permission group. \\
\hline
$permissionRequiredForRead(c)$	& returns the permission required for reading the component. \\
\hline
$permSACs(p,sac)$	& holds iff permission $p$ is required for performing the system call $sac$ (of type $SACall$). \\
\hline
\end{tabularx}
\caption{Helper functions and predicates}
\label{tab:model:semantic_descriptions}
\end{table}
\noindent The first property presents a controversial characteristic of Android's new permission system. For example, as the dangerous permissions $\mathsf{READ\_CONTACTS}$ (required for reading the contact list) and $\mathsf{WRITE\_CONTACTS}$ (required for writing the contact list) both belong to the permission group $\mathsf{CONTACTS}$, none of them can be individually granted. Instead, the application must be granted the permission group $\mathsf{CONTACTS}$, giving it the right to both reading and writing the user's contact list. 
This violates the intended least privilege security policy claimed by the designers of the platform.

\begin{prop} [No fine control over grouped permissions] \label{modelproperty1}
 \mbox{} \\
 \footnotesize
$\forall (s,s': \AndroidState) (p:\Perm) (g:\PermGrp) (app:\AppId),$ \\
$permissionIsGrouped(p) \rightarrow\ \neg \Mathexecrel{s}{\texttt{grant}~p~app/ok}{s'}$ \\ \\
\textit{Android's permission system is not granular enough for granting a proper subset of the set of permissions that belong to a group.}
\end{prop}

\noindent The next property formalizes another weak point in the specification of Android's new permission system: in a valid state an application may have the right of writing the contact list ($\mathsf{WRITE\_CONTACTS}$) even if this permission was never individually granted.

\begin{prop} [Implicit individual permission granting] \label{modelproperty2}
 \mbox{} \\
 \footnotesize
$\exists (s:\AndroidState) (p:\Perm) (app:\AppId),$
$valid\_state(s) \land \\
getPermissionLevel(p) = dangerous~ \land$ $p \notin getGrantedPermsForApp(app,s)~ \land$  \\ $p \notin getDefPermsForApp(app,s) \land appHasPermission(app,p,s)$ \\ \\
\textit{Applications may obtain privileges that were never granted to it.}
\end{prop}

In Android 6 an application that wishes to send information through the network must have the permission $\mathsf{INTERNET}$ but, since this permission is of level $normal$, any application that lists it as used in its manifest file has the right to access the network in an implicit and irrevocable way. Once again, this has been criticized due to the potential information leakage it allows. The following property formally generalizes this situation and embodies a reasonable argument to roll back this security issue introduced in Android Marshmallow.

\begin{prop}[Internet access implicitly and irrevocably allowed]
 \mbox{} \\
 \footnotesize
$ \forall (s:\AndroidState) (sac:SACall) (c:\Comp) (ic:\iComp) (p:\Perm), \\
valid\_state(s) \rightarrow permSAC(p, sac) \rightarrow \\
getPermissionLevel(p) = normal \rightarrow getPermissionId(p) \in \\
getAppRequestedPerms(getManifestForApp(getAppFromCmp(c,s),s)) \rightarrow\\
(ic, c) \in getRunningComponents(s) \rightarrow \Mathexecrel{s}{\texttt{call}~ic~sac/ok}{s}$ \\ \\
\textit{If the execution of an Android API call only requires permissions of level $normal$, it is enough for an application to list them as used on its manifest file to be allowed to perform such call.}
\end{prop}
\section{A certified reference validation mechanism}
\label{sec:excspec}
The implementation of the Android security system that we have developed consists of a set of \texttt{Coq} functions such that for every predicate involved in the axiomatic specification of action execution there exists a function which stands for the functional counterpart of that predicate. 
In this section we show how the correctness of the implementation is certified by a formal proof that establishes its soundness with respect to the inductive (axiomatic) semantics of the Android security mechanisms. The execution of an action has been implemented as a $step$ function that given a system state $s$ and an action $a$ invokes the function that implements the execution of $a$ in $s$ and returns an object $res$ of type 
$Result \eqdef \{ resp\ :\ Response, st\ :\ \AndroidState \}$, where  $res.resp$ is either an error code $ec$, if the precondition of the actions does not hold in state $s$, or otherwise the value $ok$, and the state $res.st$ represents the execution effect. 
The $step$ function acts basically as an action dispatcher. \figref{fig:step}, which shows the structure of the dispatcher, details the branch corresponding to the dispatching of action \texttt{install}, which is the action we shall use along this section to illustrate the working of the implementation.
\begin{figure}[ht]
\scriptsize
\begin{displaymath}
\begin{array}{l}
\textbf{Definition}\ step(s,a)\ :=\ \\
\quad \textbf{match}\ a\ \textbf{with} \\
\quad  \hspace{1cm} |\ \ldots \Rightarrow\ \ldots \\
\quad  \hspace{1cm} |\  \texttt{install}\ app\ m\ c\ lRes\ \Rightarrow\ install\_safe(app,m,c,lRes,s) \\
\quad  \hspace{1cm} |\ \ldots \Rightarrow\ \ldots \\
\quad\ \textbf{end}.\\  \\
\textbf{Definition}\ install\_safe(app,m,c,lRes,s)\ : Result\ :=  \\
\quad \textbf{match}\ install\_pre(app,m,c,lRes,s)\ \textbf{with} \\
\quad \hspace{1cm} |\ Some\ ec\ \Rightarrow\ \{\ap{error}{ec},s\}  \\
\quad \hspace{1cm} |\ None\ \Rightarrow\ \{ok, install\_post(app,m,c,lRes,s) \}\\
\quad\ \textbf{end}.\\
\end{array}
\end{displaymath}
\caption{The $step$ function and execution of \texttt{install} action}
\label{fig:step}
\end{figure}
\noindent The functions invoked in the branches, like $install\_safe$, are state transformers whose definition follows this pattern: first it is checked whether the precondition of the action is satisfied in state $s$, and then, if that is the case, the function that implements the execution of the action is invoked. Otherwise, the state $s$, unchanged, is returned along with an appropriate response specifying an error code which describes the failure.
In this figure we also describe the function that implements the execution of the $\texttt{install}$ action. The \texttt{Coq} code of this function, together with that of the remaining functions, can be found in \cite{AndroidCoq:2016}\footnote{We omit here the formal definition of these functions due to space constraints.}.
The function $install\_pre$ is defined as the nested validation of each of the properties of the precondition, specifying which error to throw when one of them doesn't hold.
The function $install\_post$ implements the expected behavior of the $install$ action: the identifier of the application is prepended to the list of installed applications,
both the list of granted permission and the list of granted permission groups are initialized as empty for it, its resource list is added to the system, and
its manifest, certificate and defined permissions are included in the system state\footnote{We implement the sets in the model with lists of \texttt{Coq}.}.

\subsection{ Soundness}
\label{sec:soundness}
We proceed now to outline the proof that the functional implementation of the security mechanisms of Android correctly implements the axiomatic model. This correctness property has been formally stated as a soundness theorem and verified using \texttt{Coq} \cite{AndroidCoq:2016}.

\begin{theorem} 
[Soundness of Android security system implementation] \label{theorem:soundness}
\footnotesize
 \mbox{} \\
$ \begin{array}[t]{l}
\forall\ (s:\AndroidState)\ (a:\Action),
\ap{valid\_state}{s} \rightarrow\ 
s\step{a/step(s,a).resp} \bap{step}{s}{a}.st \\
 \end{array} $
\end{theorem}

\noindent The proof of this theorem follows by, in the first place, performing a case analysis on $Pre(s,a)$ (this predicate is decidable) and then in the case that $Pre(s,a)$ applying Lemma~\ref{lemma:soundvalid}; otherwise applying Lemma~\ref{lemma:sounderror}.

\begin{lemma} 
[Soundness of valid execution] \label{lemma:soundvalid}
 \mbox{} \\
\footnotesize
$ \begin{array}[t]{l}
\forall\ (s:\AndroidState)\ (a:\Action),
\ap{valid\_state}{s} \rightarrow\ 
Pre(s, a)\ \rightarrow\\
s \step{a/ok} \bap{step}{s}{a}.st\ \wedge\  \bap{step}{s}{a}.resp=ok\\
 \end{array} $
\end{lemma}

\noindent The proof of Lemma~\ref{lemma:soundvalid} proceeds by applying functional induction on \bap{step}{s}{a} and then by providing the corresponding proof of soundness of the function that implements the execution of each action. Thus, in the case of the action \texttt{install} we have stated and proved \lemref{prop:installcorrect}. This lemma, in turn, follows by performing a case analysis on the result of applying the function $install\_pre$ on $s$ and the action: if the result is an error code then the thesis follows by contradiction. Otherwise, it follows by the correctness of the function $install\_post$.

\begin{lemma}
[Correctness of \texttt{install} execution] \label{prop:installcorrect}
\mbox{} \\
\footnotesize
$ \begin{array}[t]{l}
\forall\ (s:\AndroidState)\ (app:\App)\ (m:\Manifest)\ (c:\Cert)\ (lRes: list~\Res), \\
\ap{valid\_state}{s} \rightarrow\ 
Pre(s, \texttt{install}\ app\ m\ c\ lRes)\ \rightarrow\\
Post(s, \texttt{install}\ app\ m\ c\ lRes, install\_post(app,m,c,lRes,s))  \\
 \end{array} $
\end{lemma}

\noindent As to \lemref{lemma:sounderror}, the proof also proceeds by first applying functional induction on \bap{step}{s}{a}. Then, for each action $a$, it is shown that if $\neg Pre(s,a)$ the execution of the function that implements that action yields the values returned by the branch corresponding to the case that the function that validates the precondition of the action $a$ in state $s$ fails, i.e., an error code $ec$ and the (unchanged) state $s$.

\begin{lemma} 
[Soundness of error execution] \label{lemma:sounderror}
 \mbox{} \\
\footnotesize
$ \begin{array}[t]{l}
\forall\ (s:\AndroidState)\ (a:\Action), 
\ap{valid\_state}{s} \rightarrow\ 
\neg Pre(s,a) \rightarrow\ \exists\ (ec:ErrorCode),\\
\bap{step}{s}{a}.st = s \wedge \bap{step}{s}{a}.resp = error(ec) \wedge  ErrorMsg(s,a,ec)\\
 \end{array} $
\end{lemma}

\subsection{Reasoning over the certified reference validation mechanism}
\label{sec:secpropcert}
We have modeled the execution of the permission validation mechanism during a session of the system as a function that implements the execution of a list of actions starting in an (initial) system state. The output of the execution, a trace, is the corresponding sequence of states.
\footnotesize
\begin{displaymath}
\begin{array}{l}
	\textbf{Function}\ trace\ (s:\AndroidState)\ (actions:list\ \Action)\ :\ list\ \AndroidState \ := \\
	\quad \textbf{match}\ actions\ \textbf{with} \\
	\quad \hspace{1cm} |\ nil\ \Rightarrow\ nil \\
	\quad \hspace{1cm} |\ action::rest\ \Rightarrow\ \textbf{let}\ s'\ :=\ (step\ s\ action).st\ \textbf{in}\ s'::trace\ s'\ rest \\
	\quad \textbf{end.}
\end{array}
\end{displaymath}
\normalsize

\noindent We have stated and proved several security properties over the function $trace$.
In what follows $s$, $initstate$, $sndstate$ and $laststate$ stand for variables of type $\AndroidState$, $p$ is a variable of type $\Perm$, $app$ and $app'$ of type $\AppId$ and $l$ of type $list\ \Action$.
We present first a property that formally states that in version 6 of the OS for an application to have a non-grouped dangerous permission it must be explicitly granted to it.

\begin{prop}[Dangerous permissions must be explicitly granted]
\label{impproperty2}
 \mbox{} \\
 \footnotesize
$	valid\_state(initState) \rightarrow 
	app \in getInstalledApps(initState) \rightarrow$ \\
$	getPermissionLevel(p)= dangerous \rightarrow 
	permissionIsGrouped(p) = None \rightarrow $ \\
$	appHasPermission(app,p,lastState) \rightarrow$\\
$	\neg appHasPermission(app,p,initState)\rightarrow
	\texttt{uninstall}~app \notin l \rightarrow $ \\
$	last(trace(initState,l),initState) = lastState \rightarrow 
	\texttt{grant}~p~app \in l$ \\ \\
	\textit{A non-grouped dangerous permission can only be explicitly granted to an application.}
\end{prop}

\noindent With the following property we formally state, and prove, that if an application used to have a permission that was later revoked, only re-granting it will allow the application to have it again.
\begin{prop}[Revoked permissions must be regranted]
 \mbox{} \\
 \footnotesize
$	valid\_state(initState) \rightarrow
	getPermissionLevel(p) = dangerous \rightarrow$ \\
$	permissionIsGrouped(p) = None \rightarrow
	p \notin getDefPermsForApp(app,initState) \rightarrow$ \\
$	step(initState,\texttt{revoke}~p~app).st = sndState \rightarrow$ \\
$	step(initState, \texttt{revoke}~p~app)).resp=ok \rightarrow
	\texttt{uninstall}~app \notin l \rightarrow
	\texttt{grant}~p~app \notin l \rightarrow $\\
$	last(trace(sndState,l),sndState) = lastState \rightarrow$\\
$	\neg appHasPermission(app,p,lastState)$ \\ \\
\textit{ If an application used to have a permission that was later revoked, only regranting it will allow the application to have it again.}
\end{prop}

\noindent Certain assertions on which a developer could rely in previous versions of Android OS do not hold in its latest version. The following property states that a running component may have the right of starting another one on a certain state, but may not be able to do so at a later time. 
\begin{prop}[The right to start an external component is revocable]
 \mbox{} \\
 \footnotesize
$	\forall
	(c:\Comp) 
	(act:\Activity)$ 
$	valid\_state(initState) \rightarrow \\
	getPermissionLevel(p) = dangerous \rightarrow$ 
$	permissionIsGrouped(p) = None \rightarrow \\
	app \neq app' \rightarrow $
$	p \notin getDefPermsForApp(app,initState) \rightarrow 
	inApp(c,app,initState) \rightarrow$ \\
$	inApp(act,app',initState) \rightarrow 
	cmpProtectedByPerm(act) = Some~p \rightarrow$ \\
$	canStart(c,act,initState) \rightarrow 
	\exists (l: list\ \Action),
	\texttt{uninstall}~app \notin l\ \land$ \\
$	\texttt{uninstall}~app' \notin l \land
	\neg canStart(c,act,last(trace(initState,l), initState))$ \\ \\
\textit{A running component may have the right of starting another one on a certain state, but may not be able to do so at a later time.}
\end{prop}

When an application $app$ is granted a permission $p$ to access certain resource, it is also given the right to delegate this ability to another application, say $app'$, to access that same resource on its behalf. However, if $p$ is revoked from the application $app$, the permission delegations is not invalidated and thus the application $app'$ may still be able to access the resource. This property is a proof that the current specification allows a behavior which is arguably against the user's will. 

\begin{prop} [Delegated permissions are not recursively revoked] \label{impproperty1} 
 \mbox{} \\
 \footnotesize
$	\forall 
	(ic,ic': \iComp) 
	(c,c':\Comp)
	(u:uri)
	(cp:CProvider), $ \\
$	valid\_state(s) \rightarrow 
	step(s,\texttt{grant}~p~app).resp = ok \rightarrow $ \\
$	getAppFromCmp(c,s) = app \rightarrow 
	getAppFromCmp(c',s) = app' \rightarrow$ \\
$	(ic, c) \in getRunningComponents(s) \rightarrow
	(ic', c') \in getRunningComponents(s) \rightarrow$ \\
$	canGrant(cp,u,s) \rightarrow 
existsRes(cp,u,s) \rightarrow 
componentIsExported(cp) \rightarrow$ \\
$	permissionRequiredForRead(cp) = Some~p \rightarrow$ \\
$	\mathsf{let}~opsResult := trace(s,\texttt{[}\texttt{grant}~p~app, \texttt{grantP}~ic~cp~app'~u~Read$,\\ 
$ \texttt{revoke}~p~app\texttt{]}~\mathsf{in}$ 
$	step(last(opsResult,s), \texttt{read}~ic'~cp~u).resp=ok$ \\ \\
\textit{In Android 6 if a permission $p$ is revoked for an application $app$ not necessarily shall be revoked for the applications to which $app$ delegated $p$.} 
\end{prop}
\section{Related work}
\label{sec:relwork}
Several analyses have been carried out concerning the security of the Android system~\cite{C16,apex,forte13,felt_usenixsec2011,android2011}.
Few works, however, pay attention to the formal aspects of the permission enforcing framework. In particular,  Shin \textit{et al.} \cite{Android_formalSecurity2010,DBLP:conf/csreaSAM/ShinKFT10}
build a formal framework that represents the Android permission
system, which is developed in \texttt{Coq}, as we do. However, that work does not consider several aspects of the platform covered in our
model, namely, the different types of components, the interaction
between a running instance and the system, the R/W
operation on a content provider, the semantics of the permission
delegation mechanism and novel aspects of the security model, such as the management of runtime permissions.

Moving away from OS verification, many works have addressed the problem of relating inductively defined relations and executable
functions, in particular in the context of programming language semantics. For instance, Tollitte \emph{et al}~\cite{Tollitte:2012:CPP} show how to extract a functional implementation from an inductive specification in the \textsf{Coq}
proof assistant. Similar approaches exist for Isabelle, see e.g.~\cite{Berghofer:2009:TPHOLS}. Earlier, alternative approaches
such as~\cite{Balaa:2000:TPHOLS,Barthe:2006:FLOPS} aim to provide reasoning principles for executable specifications.
\section{Conclusion and future work}
\label{sec:conclusion}
We have presented the development of an exhaustive formal specification of the Android security model that includes elements and properties that have been partially analyzed in previous work. We have enhanced the model considered in~\cite{DBLP:conf/ictac/BetarteCLR15,BetarteCLR16} with an explicit treatment of errors and with the latest version of the security mechanisms of the platform, which make it possible to grant permissions at run time. We also present  the formaization of security properties concerning the Android permission mechanisms that have not previously been formally verified and proved. 
Using the programming language of the proof-assistant \texttt{Coq} we have defined a functional implementation of the reference validation mechanism, certified its correctness with respect to the axiomatic specification of the reference monitor and derived a certified \texttt{Haskell} prototype (\texttt{CertAndroidSec}) applying the program extraction mechanism provided by the proof assistant.
In Appendix~\ref{code}, we provide listings of part of the Haskell code that has been automatically generated using Coq. The full certified code is available in \cite{AndroidCoq:2016}. The formal development is about 21k LOC of \texttt{Coq}, including proofs, and constitutes a suitable basis for reasoning about Android's permission model and security mechanisms. 

We plan to use the certified extracted algorithm as a testing oracle and also to conduct verification activities on actual implementations of the platform. In particular, 
we are investigating the use of that algorithm to compare the results of executing an action on a real Android platform and executing that same action on the correct program. This would allow us to monitor the actions performed in a real system and assessing whether the intended security policy is actually enforced by the particular implementation of the platform. 
\newcommand{\acceso}{Access 09/17}

\newpage
\appendix
\section{Valid state} 
\label{app:validstate}
\small
The model formalizes a notion of valid state that captures several well-formedness conditions. It is formally defined as a predicate \textit{valid\_state} on the elements of type $\AndroidState$. This predicate holds on a state $s$ if the following conditions are met: 
\begin{itemize} 
\item all the components both in installed applications and in system applications have different identifiers;
\item no component belongs to two different applications present in the device;
\item no running component is an instance of a content provider;
\item every temporally delegated permission has been granted to a currently running component and over a content provider present in the system;
\item every running component belongs to an application present in the system;
\item every application that sets a value for a resource is present in the system;
\item the domains of the partial functions $\AppsManifest$, $\AppsCert$ and $\AppsDefPerms$ are exactly the identifiers of the user installed applications;
\item the domains of the partial functions $\AppsPerms$ and $\GrantedGroups$ are exatcly the identifiers of the applications in the system, both those installed by the users and the system applications;
\item every installed application has an identifier different to those of the system applications, whose identifiers differ as well;
\item all the permissions defined by applications have different identifiers;
\item every partial function is indeed a function, that is, their domains don't have repeated elements;
\item every individually granted permission is present in the system; and
\item all the sent intents have different identifiers.
\end{itemize}

All these safety properties have a straightforward interpretation in our model. We omit here the formal definition of \textit{valid\_state} due to space constraints. The full formal definition of the predicate is available in \cite{AndroidCoq:2016}. 
\normalsize

\section{Generated code} \label{code}
Just for the sake of illustration, in what follows we provide listings of part of the Haskell code that has been automatically generated using the Coq extraction mechanism. The code is annotated with inline comments and manually indented to fit on the page width. 

We have included the definition of the System, as a datatype and  the complete definition of the action \texttt{install} and the code of the dispatcher,  which implements the execution of an action in a given state.

\newpage

\begin{lstlisting}[frame=lines,basicstyle=\tiny,caption={The System}, label={lst:system-state}]
{- The System is represented as a datatype
 - comprising a State and an Environment 
 -}

data System =
   Sys State Environment

-- The Environment datatype
data Environment =
   Env
   -- The manifest and certificate of installed user applications 
     (Mapping IdApp Manifest)   
     (Mapping IdApp Cert)
   -- The permissions defined by the applications
     (Mapping IdApp (([]) Perm0))  
   -- System applications
     (([]) SysImgApp)                      

-- The datatype State
data State =
   St 
   -- The installed user applications 
     (([]) IdApp)  
   -- Granted group and individual permissions for each application 
     (Mapping IdApp (([]) IdGrp)) 
     (Mapping IdApp (([]) Perm0))
   -- Running components and their instances
     (Mapping ICmp Cmp)    
   -- Permanent and temporary permission delegations
     (Mapping ((,) ((,) IdApp CProvider) Uri) PType)  
     (Mapping ((,) ((,) ICmp CProvider) Uri) PType)
   -- Values of resources
     (Mapping ((,) IdApp Res) Val) 
   -- Sent intents
     (([]) ((,) ICmp Intent0)) 
\end{lstlisting}

\begin{lstlisting}[frame=lines,basicstyle=\tiny,caption={Install checks}, label={lst:installprec}]
{- Install semantics:
 - install_pre checks whether an installation can take place in a state.
 - It returns the corresponding ErrorCode if the installation is not allowed
 -}

install_pre :: IdApp -> Manifest -> Cert -> (([]) Res) ->
    System -> Prelude.Maybe ErrorCode
install_pre app0 m c lRes s =
-- The application can not be already installed
  case isAppInstalledBool app0 s of {
   Prelude.True -> Prelude.Just App_already_installed;
   Prelude.False ->
-- Components in the application must have different identifiers
    case has_duplicates idCmp_eq (map getCmpId (cmp m)) of {
     Prelude.True -> Prelude.Just Duplicated_cmp_id;
     Prelude.False ->
-- The defined permissions must have different identifiers.
      case has_duplicates idPerm_eq (map idP (usrP m)) of {
       Prelude.True -> Prelude.Just Duplicated_perm_id;
       Prelude.False ->
-- The new components' ids must differ from those already present in the system.
        case existsb (\c0 -> cmpIdInStateBool c0 s) (cmp m)
        of {
         Prelude.True -> Prelude.Just Cmp_already_defined;
         Prelude.False ->
-- No permission defined by other application can be redefined.
          case Prelude.not (authPermsBool m s) of {
           Prelude.True -> Prelude.Just
           Perm_already_defined;
           Prelude.False ->
-- All the intent filters must be well defined.
            case anyDefinesIntentFilterIncorrectly (cmp m)
            of {
             Prelude.True -> Prelude.Just
                Faulty_intent_filter;
-- If everything is ok, then no error is returned.
             Prelude.False -> Prelude.Nothing}}}}}}
\end{lstlisting}

\newpage

\begin{lstlisting}[frame=lines,basicstyle=\tiny,caption={Install effect}, label={lst:installpost}]
{- The function install_post compute the state resulting from 
 - installing a fresh application
 -}

install_post :: IdApp -> Manifest -> Cert -> (([]) Res) ->
    System -> System
install_post app0 m c lRes s =
  let {oldstate = state s} in
  let {oldenv = environment s} in
-- The application identifier is appended to the list of installed applications
  Sys (St ((:) app0 (apps oldstate))
-- with no permission or permission groups granted,
  (map_add idApp_eq (grantedPermGroups oldstate) app0 ([]))
  (map_add idApp_eq (perms oldstate) app0 ([])) (running
  oldstate)
  (delPPerms oldstate) (delTPerms oldstate)
-- its resources are initialized with the default value, and
  (addNewResCont app0 (resCont oldstate) lRes) (sentIntents
  oldstate))
-- its manifest, certificate and permissions are stored in the state.
  (Env
  (map_add idApp_eq (manifest oldenv) app0 m)
  (map_add idApp_eq (cert oldenv) app0 c)
  (map_add idApp_eq (defPerms oldenv) app0 (nonSystemUsrP m))
  (systemImage oldenv))
\end{lstlisting}

\begin{lstlisting}[frame=lines,basicstyle=\tiny,caption={Safe install}, label={lst:safeinstall}]
{- The function install_safe checks for errors using the function install_pre,
 - returning the state unmodified along with the error code, if there is one.  
 - Otherwise, it computes the new state by executing install_post
 -}

install_safe :: IdApp -> Manifest -> Cert -> (([]) Res) -> System -> Result0
install_safe app0 m c lRes s =
  case install_pre app0 m c lRes s of {
   Prelude.Just ec -> Result (Error0 ec) s;
   Prelude.Nothing -> Result Ok (install_post app0 m c lRess)}
\end{lstlisting}

\begin{lstlisting}[frame=lines,basicstyle=\tiny,caption={Dispatcher}, label={lst:dispatcher}]
{- The function step is just a dispatcher which
 - performs pattern matching on the action to be
 - executed and calls the corresponding function
 - (for example, install_safe)
 -}

step :: System -> Action -> Result0
step s a =
  case a of {
   Install app0 m c lRes -> install_safe app0 m c lRes s;
   Uninstall app0 -> uninstall_safe app0 s;
   Grant p app0 -> grant_safe p app0 s;
   Revoke p app0 -> revoke_safe p app0 s;
   GrantPermGroup grp app0 -> grantgroup_safe grp app0 s;
   RevokePermGroup grp app0 -> revokegroup_safe grp app0 s;
   HasPermission a0 p -> Result Ok s;
   Read0 ic cp u -> read_safe ic cp u s;
   Write0 ic cp u v -> write_safe ic cp u v s;
   StartActivity intt ic -> startActivity_safe intt ic s;
   StartActivityForResult intt n ic -> startActivity_safe intt ic s;
   StartService intt ic -> startService_safe intt ic s;
   SendBroadcast intt ic p -> sendBroadcast_safe intt ic p s;
   SendOrderedBroadcast intt ic p -> sendBroadcast_safe intt ic p s;
   SendStickyBroadcast intt ic -> sendStickyBroadcast_safe intt ic s;
   ResolveIntent intt a0 -> resolveIntent_safe intt a0 s;
   ReceiveIntent intt ic a0 -> receiveIntent_safe intt ic a0 s;
   Stop ic -> stop_safe ic s;
   GrantP ic cp a0 u pt -> grantP_safe ic cp a0 u pt s;
   RevokeDel ic cp u pt -> revokeDel_safe ic cp u pt s;
   Call ic sac -> call_safe ic sac s }
\end{lstlisting}
\end{document}